\documentclass[preprintnumbers,amsmath,amssymb,twocolumn]{revtex4}
\usepackage[dvips,colorlinks=true,citecolor=red,filecolor=green,linkcolor=blue,pdfnewwindow=true]{hyperref}
\usepackage{graphicx}
\usepackage[title]{appendix}

\begin{document}
\title{Power law nature in electron solid interaction}
\author{Moirangthem Shubhakanta Singh$^{1}$ and R.K. Brojen Singh$^{2}$}
\email{brojen@jnu.ac.in (Corresponding author)}
\affiliation{$^1$Department of Physics, Manipur University,Canchipur-795003, Manipur, India.\\
$^2$School of Computational and Integrative Sciences, Jawaharlal Nehru University, New Delhi-110067, India.}

\begin{abstract}
Monte carlo simulation of paths of a large number of impinging electrons in a multi-layered solid allows to define area of spreading electrons (A) to capture overall behavior of the solid. This parameter 'A' follows power law with electron energy. Further, change in critical energies, which are minimum energies lost corresponding to various electrons, as a function of variation in lateral distance also follows power law nature. This power law behavior could be an indicator of how strong self-organization a solid has which may be used in monitoring efficiency of device fabrication.
\end{abstract}


\maketitle

\section{Introduction}
The interaction of beam of energetic electrons with the target solid material technique, which is electron beam lithography, is of great interest in probing material properties (chemical, electrical, physical etc) \cite{kyser} at sub-micron and nanoscale level \cite{zhou}, and has many applications, modeling radio-induced cellular damages \cite{shi}, in surface science technology \cite{nie}, nanolithography techniques \cite{gri}, fabrication of fractal surfaces \cite{sto}, various biological applications \cite{cet} etc. The impinging energetic electron suffers random collisions from a number of scattering centres with random distribution of potentials in the centres in the solid materials of various layers \cite{shi}, and the electron follows a stochastic path inside the solid material \cite{kyser}. The analysis of the electron paths could highlight some of the important properties of the material which will be used in various device fabrications and various other applications.

One of the most important properties of real networks, ranging from social to biological protein-protein networks \cite{bara}, is power law nature of the network distribution \cite{bara1} which could be a reflection of fractal nature of the system \cite{jung}. Since fractal behavior of the system can be used as an indicator of self-organization in the system \cite{kauf}, one can use this property to identify important patterns and their origin in the system \cite{mand,peit,thei}. The path of the penetrating electrons in solid system is the reflection of organization of the regular scattering centres with random potential distributions, one can identify probable parameters to capture patterns of organization in the solid material. In this work, we try to search for possible parameters to characterize fractal nature of the solid systems using Monte carlo simulation procedure which could be used for various fabrication techniques. In the section II the detailed Monte carlo procedure of electron-solid interaction is described. Simulation results are described in section III, and some conclusions are drawn based on the simulation results.

\section{Electron solid interaction model}
The path traversed by impinging energetic electrons in solid is based on the electron transport within the stochastic formalism of scattering process of electrons with the solid along their trajectory \cite{shu,shi}. The penetrating electrons encounter randomly distributed scattering centres within the electron interaction range \cite{shi}, and the electrons undergo complicated brownian paths inside the solid \cite{mur}. These electrons with energy ($E$) move in straight lines between any two scattering centres, and once they suffer interaction with scattering centres of the solid, the change in their directions are defined by $(E,\theta,\phi)$, where, $\theta$ and $\phi$ are scattering and azimuthal angles respectively. The solids could be single or multi-layered thin film with different distributions of scattering centres in different solid layers.
\begin{figure*}
\label{sl}
\begin{center}
\includegraphics[height=11cm,width=13.5cm]{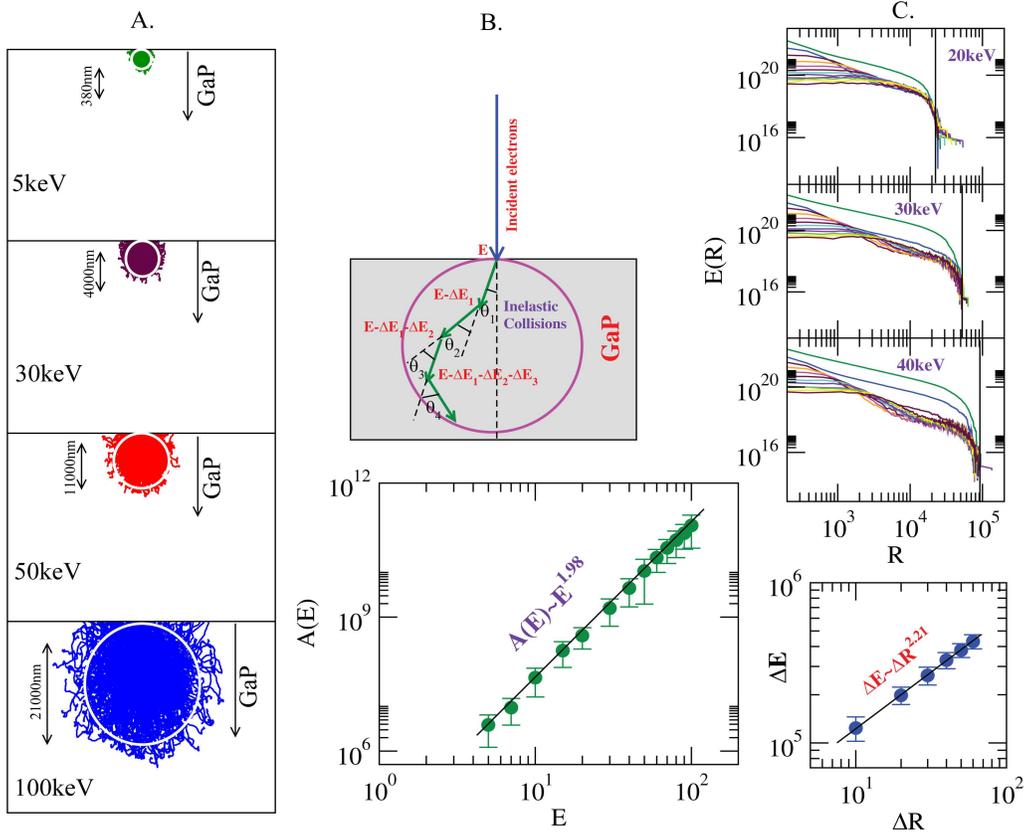}
\caption{The plots in left panel show the trajectories of 500 electrons in single solid $GaP$ thin film with area of spreading of electron paths (circles) for different energies 5keV, 30keV, 50keV and 100keV. The schematic diagram of impinging electrons with random scattering events are shown in upper middle panel. The plot of $A$ versus $E$ is shown in middle lower panel. Solid line is the fit on the data. The right panels show the plots of loss in energy as a function of lateral distance $R$ for various energies of the electrons. The lower right panel show the plot of $\Delta E$ versus $\Delta R$ and solid line is the fit on the data.} 
\end{center}
\end{figure*}

We consider impinging electrons suffer elastic collisions from scattering centres distributed in the single or multi-layered thin solid film, where, differential cross section can be described by classical screened Rutherford's formula \cite{mur},
\begin{eqnarray}
\label{cross}
\frac{d\sigma}{d\Omega}=\frac{e^4Z(Z+1)}{4E[1+cos(\theta)+2\beta]^2}
\end{eqnarray}
where, $e$ is electronic charge, $Z$ is the atomic number of the material and $d\Omega$ is solid angle. $\beta$ is screening parameter to account for electrostatic screening of the nucleus by the orbital electrons, and is given by Thomas-Fermi model of the atomic field \cite{nig},
\begin{eqnarray}
\beta=0.316\left(\frac{\hbar Z^{1/3}}{pa_o}\right)
\end{eqnarray}
where, $a_o$ is Bohr radius, $\hbar$ is Planck's constant and $p$ is the electron's momentum. The scattering angle $\theta$ can be calculated by evaluating total elastic cross section $\sigma=\int d\Omega$ from equation (\ref{cross}),
\begin{eqnarray}
\label{theta}
\theta=cos^{-1}\left[1+\frac{2\beta F(\theta)}{1+\beta-F(\theta)}\right]
\end{eqnarray}
where, $F(\theta)$ is accumulated function of scattering probability \cite{mur} which is a function of $theta$ only. In the Monte carlo simulation procedure, $\theta$ can be obtained by generating a uniform random number $R_1$ in $[0,1]$, and azimuthal angle $\phi$ by generating another uniform random number $R_2$ and using,
\begin{eqnarray}
\label{phi}
\phi=2\pi R_2
\end{eqnarray}
Thus, $\theta$ and $\phi$ can be calculated by generating two sets of uniform random numbers, and using equation (\ref{theta}) and (\ref{phi}).
 
\subsection{Energy loss calculation of traversing electron}
The energy of the electron, suffering interaction from the scattering centres distributed randomly along its path, continuously looses its kinetic energy and can be calculated using Behte's continuous slowing down approximation model \cite{bethe}. This model is a good emperical method for high energetic electrons as compared to ionization energy $J$ i.e. for $E>>J$, but suffers problem for $E\le J$ \cite{rao}. However, the model was generalized for all range of energies \cite{joy}, where, the energy loss $\Delta E$ of the penetrating electron a path length $L$ along its trajectory is given by,
\begin{eqnarray}
\label{DE}
\Delta E&=&-\int_0^L dz\left(\frac{dE}{dz}\right);\\
&&\frac{dE}{dz}=-2\pi e^4\left(\frac{\rho N_A}{ME}\right)ln\left(\frac{1.166E}{\epsilon}\right)\nonumber\\
&&\epsilon=\frac{J}{1+C(J/E)}\nonumber
\end{eqnarray}
where, $M$ is the atomic weight of the target material, $\rho$ is the density, $N_A$ is the Avogadro's number and $C$ is a constant ($C\rightarrow 1;~C<1$). The mean ionization energy $J$ can be obtained from the emperical formula \cite{ber},
\begin{eqnarray}
\label{J}
\frac{J}{Z}=9.76+58.8Z^{-1.19}
\end{eqnarray}
where, $\frac{J}{Z}\rightarrow 9.76$ as $Z\rightarrow\infty$. Further, the sensitivity of the $J$ in Monte carlo simulation can be controlled by taking logarithm of this parameter.

\subsection{Modeling electron path in multi-layered system}
The mean free path for single layered system, calculated using equation (\ref{cross}), can be extended for multi-layered system by defining a probability $P_m(u)$ that the electron once scattered from first layer is not scattered until $m$th layer \cite{hori,haw}, and can be obtained from the following equation,
\begin{eqnarray}
\label{prob}
\frac{dP_m(u)}{du}=-\Gamma_mP_m(u)
\end{eqnarray}
where, $\Gamma_m$ is scattering probability per unit length in $m$th layer. Boundary condition is taken as $P_1(0)=1$. Solving equation (\ref{prob}) one can arrive at $P_{m+1}(u-u_m)=P_m(u)$, where, $u_m$ is the distance between $m$th and $(m+1)$th layers along z-axis. This $P_m(u)$ can be related to $F(u)$ by, $F(u)=1-P_m(u)=1-R_1$, and $\theta$ can be solved using equation (\ref{theta}). The mean free path for single layer system is calculated as $\lambda_1=\int_{0}^{\infty}uP_1(u)du=-1/\Gamma_1$, where, $P_1(u)=exp(-u\Gamma_1)$. Similarly, mean free path for two layered system is given by, $\lambda_2=\int_{0}^{u}u^\prime P_1(u^\prime)du^\prime+\int_{u}^{\infty}u^\prime P_1(u^\prime)du^\prime$$=1/\Gamma_1+(1/\Gamma_2-1/\Gamma_1)exp(u\Gamma_1)$. Proceeding in the same way, mean free path for $m$ layered system can be calculated using,
\begin{eqnarray}
\label{m}
\lambda_m&=&\int_{0}^{u_1}u^\prime P_1(u^\prime)du^\prime+\int_{u_1}^{u_2}u^\prime P_1(u^\prime)du^\prime+\nonumber\\
&&\dots+\int_{u_m}^{\infty}u^\prime P_1(u^\prime)du^\prime
\end{eqnarray}
From equation (\ref{m}) one can able to calculate $u_m$ of impinging electron in $m$th layered material system. Now, starting from an initial vector $(x_0,y_0,z_0)^T$, we can trace the path of the penetrating electron in $m$-layered system using the following recursive procedure,
\begin{eqnarray}
\label{path}
\left(\begin{array}{c}x_{n+1}\\y_{n+1}\\z_{n+1}\end{array}\right)=\left(\begin{array}{c}x_{n}\\y_{n}\\z_{n}\end{array}\right)+u_n\left(\begin{array}{c}sin\theta_ncos\phi_n\\sin\theta_ncos\phi_n\\cos\theta_n\end{array}\right)
\end{eqnarray}
Thus the path of the penetrating electron in $m$-layered material system can be traced using the recursive procedure in (\ref{path}) with average energy loss $\Delta E$ within the Monte carlo simulation.
\begin{figure*}
\label{db}
\begin{center}
\includegraphics[height=10.5cm,width=14cm]{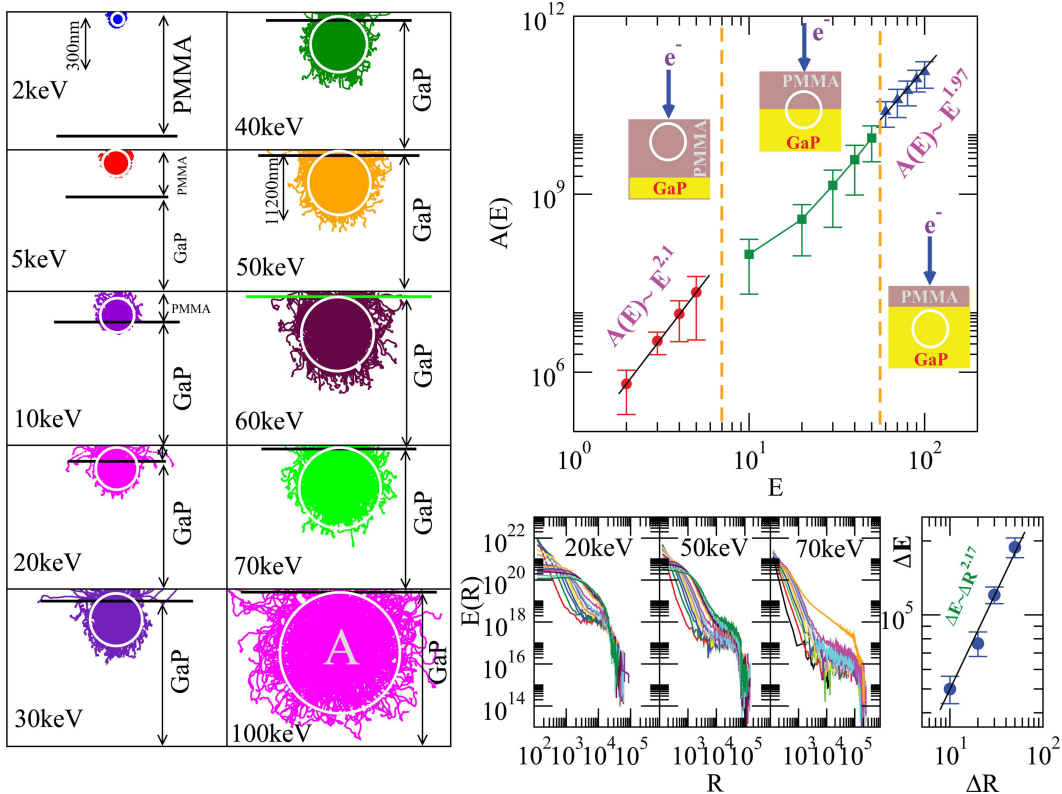}
\caption{The plots in the two columns in the left panel show the trajectories of 500 electrons in single solid double layered system of $GaP$ with resist $PMMA$ thin film with area of spreading of electron paths (circles) for different energies [2-100]keV. The plot of $A$ versus $E$ is shown in the upper right panel with different interaction regimes. Solid line is the fit on the data. The lower panels show the plots of loss in energy as a function of lateral distance $R$ for various energies of the electrons. The lower right panel show the plot of $\Delta E$ versus $\Delta R$ and solid line is the fit on the data.} 
\end{center}
\end{figure*}

\section{Results and discussion}
We consider single layered $GaP$ and double layered system of $GaP$ with resist $PMMA$ ($C_5H_8O_2$), and simulated using the Monte carlo simulation procedure described in the previous section to trace the trajectory of impinging energetic electrons in the systems and energy loss. The initial positions of the penetrating 500 electrons are taken as the same as $(x_0,y_0,z_0)^T=(0,0,0)^T$ for a range of energy [2-100]keV. The trajectory of each electron of energy $E$ is calculated using Monte carlo procedure (\ref{path}), and energy loss ($\Delta E$) during the process of penetration is obtained using equation (\ref{DE}). In our simulation the ionization energies of $C$, $H$ and $O$ are taken to be 78eV, 18.7eV and 89eV, and for $GaP$, equation (\ref{J}) is used to calculate its $J$. Since the ionization energy of both $GaP$ and $PMMA$ are of the order of eV, $E>>J$ in our case and therefore, we take $C<1$ in equation (\ref{DE}) during the calculation. The thickness of the $PMMA$ resist is taken to be $10^{-6}$metre in the double layered calculation.

\subsection{Power law nature of electron spreading area}
The paths of the impinging electrons (500 electrons' paths) in single layer $GaP$ system with different energies ([2-100]keV) are calculated (Fig. 1 left panels) and two dimensional areas of the spreading electrons (circles in the figures) for different energies are obtained. The area of each circle is calculated for ten ensembles, and average of minimum and maximum areas bounded to the two dimensional electron spreading area of 500 electron trajectories (error bars in the plot in middle lower panel of Fig. 1). This calculated areas $A$ as a function of $E$ is found to follow the following power law behaviour,
\begin{eqnarray}
\label{AE}
A(E)\sim E^\gamma
\end{eqnarray}
The straight line is the fitting curve to the calculated data. The value of $\gamma$ is found to be 1.98.

We then calculated the $A$s in double layered material system ($1\mu$m thickness for $PMMA$ and rest for $GaP$) for various electron energies [2-100]keV (Fig. 2 left panels). The behavior of $A$ with respect to $E$ (Fig. 2 middle upper panel) has three regimes, left (electrons paths are within single first layer only) and right (dominated by second layer as compared to first layer) regimes follow similar power law given by equation (\ref{AE}), and the values of $\gamma$ are found to be 2.1 and 1.97 respectively. The middle regime, which is due to contributions from both first and second layers, and does not follow exact power law nature.

\subsection{Power law behaviour in energy loss}
The energy loss of impinging 1500 electrons in single layer system are calculated as a function of lateral distance $R$ of the electron trajectories for different energies (Fig. 1 right panels). The values of $E$ sharply drop after a certain value of $R$ for different values of $E$ showing that the electrons do not have sufficient energies to penetrate further in the solid. We then calculated these critical $E_c$ and $R_c$ for different $E$s in the range [10-100]keV. Then we calculated possible changes in these critical energies ($\Delta E$) as a function of $\Delta R$ starting from lowest $(E_c,R_c)$, and error bars are standard deviations of the thicknesses of the drop curves (lines drawn parallel to $E$ axis in the Fig. 1 right panels). The calculated $\Delta E$ again follows power law with $\Delta R$ as follows,
\begin{eqnarray}
\label{DER}
\Delta E\sim \Delta R^\delta
\end{eqnarray}
The power law fit to the data gives the power law exponent to be $\delta=2.21$.

We now calculated $\Delta E$ in double layered material system as a function of $\Delta R$ for energies [10-100]keV (Fig. 2 left lower panels). Surprisingly, even though there is contributions from first and second layers, the behavior of $\Delta E$ as a function of $\Delta R$ follows the similar power law nature as equation (\ref{DER}) with the value of $\delta=2.17$.

\section{Conclusion}
The path and spread of the trajectory of any energetic electron is proportional to the energy it posseses and material in which the electron is penetrating. Even though the impinging electrons trajectories are stochastic zig-zag nature, the overall behavior of the large number of electrons exhibit the nature of the material's characteristics. The area of the impinging electrons in single layered system follows power law as a function of electrons' energy.

The power law behavior is found to various systems, starting from social systems to brain protein-protein networks which indicate important functional and organizational characteristics. This behavior indicates that the properties of the system is independent of scale of the system. Since this law also reflects the fractal nature of the system, it characterizes as an indicator of self-organized behavior of the material system. The impinging energetic electrons experience the self-organized behavior of the system which is reflected in the parameters calculated using Monte carlo simulation procedure. This idea of fractal nature could be used as an order parameter in the fabrication of multi-layered device with proper efficiency.

\end{document}